\documentclass[12pt]{article}
\begin{document}
\renewcommand{\thetable}{\Roman{table}}
\def \beq{\begin{equation}}
\def \bk{\bar K^0}
\def \ddr{\ddot{\bf r}}
\def \dr{\dot{\bf r}}
\def \dvx{\dot{\bf x}}
\def \dx{\dot x}
\def \dy{\dot y}
\def \dz{\dot z}
\def \eeq{\end{equation}}
\def \ev{\left[ \begin{array}{c} x_0 \\ y_0 \end{array} \right] }
\def \evl{\left[ \begin{array}{c} 1 \\ -i \end{array} \right]}
\def \evL{\left[ \begin{array}{c} \epsilon \\ 1 \end{array} \right]}
\def \evr{\left[ \begin{array}{c} 1 \\ i \end{array} \right]}
\def \evS{\left[ \begin{array}{c} 1 \\ \epsilon \end{array} \right]}
\def \k{K^0}
\def \L{|L \rangle}
\def \m{{\cal M}}
\def \n{{\cal N}}
\def \ok{\overline{K}^0}
\def \Ome{{\Omega_{\rm E}}}
\def \R{|R\rangle}
\def \s{\sqrt{2}}
\def \te{\tilde{\epsilon}}
\def \vf{{\bf F}}
\def \vo{{\bf \Omega}_{\rm E}}
\def \vr{{\bf r}}
\def \vx{{\bf x}}
\rightline{EFI-99-51}
\rightline{hep-ph/9912506}
\vspace{0.5in}
\centerline{\bf CLASSICAL ILLUSTRATIONS}
\centerline{\bf OF CP VIOLATION IN KAON DECAYS
\footnote{To be submitted to Am.~J.~Phys.}}
\vspace{0.5in}
\centerline{\it Jonathan L. Rosner}
\centerline{and}
\centerline{\it Scott A. Slezak}
\bigskip
\centerline{\it Enrico Fermi Institute and Department of Physics}
\centerline{\it University of Chicago, Chicago, IL 60637}
\bigskip

\centerline{\bf ABSTRACT}
\medskip
\begin{quote}

It is easy to construct classical 2-state systems illustrating the
behavior of the short-lived and long-lived neutral $K$ mesons in the limit
of CP conservation. The emulation of CP violation is more tricky, but is
provided by the two-dimensional motion of a Foucault pendulum. Analogies
are drawn between the pendulum and observables in neutral kaon decays.  
An emulation of CP- and CPT-violation using electric circuits is also
discussed.

\end{quote}
\bigskip

\centerline{\bf I.  INTRODUCTION}
\bigskip
Two-state systems abound in quantum mechanics, and have the virtue of
permitting exact solutions \cite{Ramsey}.  One such system is that of the
neutral kaons $\k$ and $\bk$ \cite{Baym}, which mix with one
another \cite{GP} to form short-lived and long-lived mass eigenstates.
Previously \cite{TTTV} we have shown that one can emulate this system
using coupled resonant circuits, with the violation of CP invariance
\cite{CCFT} exhibited in kaon decays corresponding to asymmetric coupling
between the circuits.

We sought to implement the suggestion of Ref.~\cite{TTTV} in a physical device,
such as a pair of coupled pendula which may be used to illustrate the
CP-conserving limit \cite{BW}.  In the course of this activity,
we came upon a familiar
system which has many of the features of the neutral kaon system, including
the asymmetric coupling between two oscillation modes which leads to the
phenomenon of CP violation.  This system is the Foucault pendulum.  In
the present article we explore the parallels between the Foucault pendulum
and the neutral kaon system, showing how one might construct a pendulum
with the closest possible relation to the actual short-lived and long-lived
neutral kaon states, and discussing not only CP, but T and CPT violation
as well.

This article is arranged as follows.  We first recall the motion of the
pendulum in Section II, review the neutral kaons in Section III (partly a
recapitulation of results from Ref.~\cite{TTTV}), and explore the parallels
in Section IV.  Some suggestions for modifying the basic pendulum to
make it emulate the actual kaon system are made in Section V, while
Section VI concludes.  An Appendix deals with the CPT-violating case.
\bigskip

\centerline{\bf II.  FOUCAULT PENDULUM}
\bigskip

\leftline{\bf A.  Equations of motion in a rotating frame}
\bigskip

The motion of a body subject to a force $\vf$ in a system rotating with
constant angular velocity $\vo$ is described by \cite{HG}
\beq \label{eqn:Cor}
m \ddr = \vf - 2m (\vo \times \dr) - m \vo \times (\vo
\times \vr) - 2 m \beta \dr~~~,
\eeq
where the second term is the Coriolis force, the third term is the centripetal
acceleration, and the last term has been inserted to describe damping due to
air resistance.  For the case in question, $\vo$ will be a vector pointing
toward the Earth's North Pole, with magnitude $2 \pi$ d$^{-1}$.

Consider a pendulum with a spherically symmetric support point so that it
is free to move in two directions.  Denote the corresponding axes by
$\hat x$ (East), $\hat y$ (North), and $\hat z$ (up, i.e., perpendicular to
the surface of the Earth).  The components of $\vo$ are
\beq
\Ome_x = 0~~,~~~\Ome_y = \Ome \cos \theta~~,~~~
\Ome_z = \Ome \sin \theta~~~,
\eeq
where $\theta$ is the latitude (positive for North latitude).
The components of $\vo \times \dr$ are
\beq
(\vo \times \dr)_x = - \Ome \sin \theta \dy~~,~~~
(\vo \times \dr)_y = \Ome \sin \theta \dx~~,~~~
(\vo \times \dr)_z = -\Ome \cos \theta \dx~~~,
\eeq
where we have neglected $\dz$ for small oscillations of the pendulum.

The centripetal acceleration term  $- m \vo \times (\vo \times r)$ in
Eq.~(\ref{eqn:Cor}) has magnitude $m \Ome^2 a \cos \theta$ and acts in
the direction $- \hat z \cos \theta + \hat y \sin \theta$, where $a$
is the radius of the Earth and $\Ome^2 a = 3.38$ cm s$^{-2}$ \cite{HG}.
It thus changes the local acceleration of gravity slightly, $g \to g_{\rm
eff}$, and displaces the equilibrium position of the pendulum.  We shall
take account of these effects by redefining $g \equiv g_{\rm eff}$ and
setting $x = y = 0$ to be the equilibrium position. 

We then define $\omega_0^2 \equiv g/l$, where $l$ is the length of the
pendulum, and $\Omega \equiv \Ome \sin \theta$,
write Eq.~(\ref{eqn:Cor}) in component form, and cancel a factor of $m$.
The result is
\beq \label{eqn:coupled}
\ddot{x} = - \omega_0^2 x + 2 \Omega \dy - \beta \dx~~,~~~
\ddot{y} = - \omega_0^2 y - 2 \Omega \dx - \beta \dy~~~.
\eeq 

The coupled equations (\ref{eqn:coupled}) can be solved for periodic motion
by substituting $x = x_0 \exp(-i \omega t)$, $y = y_0 \exp(-i \omega t)$,
leading to an eigenvalue equation for $\omega^2$.  Expanding around
$\omega = \omega_0$ and dividing by $2 \omega_0$, the result is
\beq
\m \ev = \omega \ev~~~,
\eeq
where
\beq \label{eqn:mass}
\m \equiv \left[ \begin{array}{c c} \omega_0 - i \beta & -i \Omega \\
i \Omega & \omega_0 - i \beta \end{array} \right]~~~.
\eeq
Eq.~(\ref{eqn:mass}) is very close to the result which one would obtain for
mixing of a neutral-kaon system in which the $\k$ is represented by the
basis vector $\hat x$ while the $\bk$ is represented by the basis
vector $\hat y$.  We shall explore this parallel presently.  First,
however, we show that Eq.~(\ref{eqn:mass}) leads to the familiar behavior of
the Foucault pendulum in which the plane of linear oscillations precesses
by a daily amount which depends on the latitude.
\bigskip

\leftline{\bf B.  Solution of equations of motion}
\bigskip

The normalized eigenmodes of the system (\ref{eqn:mass}) are
\beq \label{eqn:modes}
\R \equiv \frac{1}{\s} \evr~~,~~~\L \equiv \frac{1}{\s} \evl~~~,
\eeq
with eigenvalues
\beq
\mu_R = \omega_0 - i \beta + \Omega~~,~~~
\mu_L = \omega_0 - i \beta - \Omega~~~.
\eeq
An arbitrary two-component solution $\vx(t)$ describing motion in the $x-y$
plane can then be written as a linear combination of $\R$ and $\L$ as
\beq \label{eqn:sol}
\vx(t) = c_R \R e^{-i \mu_R t} + c_L \L e^{-i \mu_L t}~~~.
\eeq
The initial conditions on $\vx(0)$ and $\dvx(0)$ then permit us to
specify the complex quantities $c_R$ and $c_L$, by imposing the condition
\beq
{\rm Re}~\vx(0) = \left[ \begin{array}{c} x(0) \\ y(0) \end{array}\right]~~,~~~
{\rm Re}~\dvx(0) = \left[ \begin{array}{c} \dot{x}(0) \\ \dot{y}(0)
 \end{array} \right]
\eeq
on the solution (\ref{eqn:sol}) at $t=0$.

Let us now assume the pendulum is initially displaced in the $\hat x$ (East)
direction, with zero velocity, so that $x(0) = x_0$, $\dot{x}(0) = y(0)
= \dot{y}(0) = 0$. We then find 
\beq
{\rm Re}~c_R = \frac{x_0}{\s} \frac{\omega_0 - \Omega}{\omega_0}~~,~~~
{\rm Re}~c_L = \frac{x_0}{\s} \frac{\omega_0 + \Omega}{\omega_0}~~~,
\eeq
\beq
{\rm Im}~c_R = {\rm Im}~c_L = \frac{\beta x_0}{\s \omega_0}~~~.
\eeq
The solution for the motion of the pendulum as a function of time is
$$
x(t) = x_0 e^{- \beta t} \left[ (\cos \omega_0 t + \frac{\beta}{\omega_0}
\sin \omega_0 t) \cos \Omega t + \frac{\Omega}{\omega_0} \sin \omega_0 t
\sin \Omega t \right]~~~,
$$
\beq \label {eqn:mot}
y(t) = x_0 e^{- \beta t} \left[ (\cos \omega_0 t + \frac{\beta}{\omega_0}
\sin \omega_0 t) \sin \Omega t - \frac{\Omega}{\omega_0} \sin \omega_0 t
\cos \Omega t \right]~~~.
\eeq
For $\Omega \ll \omega_0$, one sees by making the replacement $\Omega t
\to \Omega t + \frac{\pi}{2}$ and comparing the terms $\sim \sin \Omega t$ and
$\cos \Omega t$ in Eq.~(\ref{eqn:mot}) that the plane of oscillation rotates
with angular frequency $-\Omega = -\Ome \sin \theta$.
For $\theta = 0$ there is no precession of the plane since
at the Equator the only effect of the Earth's rotation to lowest order is
a change in the effective value of $g$.  At the North or South pole the
pendulum oscillates in a fixed plane in an inertial frame, and thus its
plane of oscillation rotates with respect to the frame fixed with respect to
the Earth with angular velocity $\mp \Ome$.
\bigskip

\centerline{\bf III.  THE TWO-STATE KAON SYSTEM}
\bigskip

The neutral kaon $K^0$ was first identified in cosmic radiation in the late
1940's \cite{RB}, via its decay to a pair of charged
pions $\pi^+ \pi^-$ in a cloud chamber.  It was
``strange'' because its production occurred much more rapidly than
its decay.  Gell-Mann and Nishijima \cite{GN} explained this feature by
assigning the $K^0$ a ``strangeness'' quantum number $S = 1$, 
conserved in production processes but violated in decays.  A typical production
process, for example, would be
\beq
\pi^-~(S = 0) + p~(S = 0) \to K^0 (S=1) + \Lambda (S=-1)~~~,
\eeq
where $\Lambda$ is another ``strange'' particle first observed in the late
1940's.  Strong interactions were assumed to conserve strangeness, while weak
interactions (such as those responsible for $K^0 \to \pi \pi$) could
violate it by a maximum of one unit.

The strangeness scheme implied that the $K^0$ could not be its own antiparticle
(in contrast to some other neutral particles like the photon and the neutral
pion $\pi^0$), since it carried an additive quantum number $S = 1$.  There then
had to exist a neutral kaon with $S = -1$, the $\ok$.  When Gell-Mann proposed
this scheme, Enrico Fermi asked him what would distinguish the
$K^0$ from the $\ok$, since they would have equal masses and each could decay
to $\pi \pi$.  The solution, proposed by Gell-Mann and Pais, \cite{GP}, was
that one linear combination of the $K^0$ and $\ok$, namely $K_1 \equiv (K^0 +
\ok)/\s$, would be able to decay to $\pi \pi$, while the orthogonal
combination, $K_2 \equiv (K^0 - \ok)/\s$, would be forbidden from decaying
to $\pi \pi$ and thus would be longer-lived.  This particle was searched for
and found \cite{KL}.  Its main decay modes are $3 \pi$, $\pi e \nu$, and
$\pi \mu \nu$.  These three-body decays have smaller phase space and thus
occur with a smaller rate than the $2 \pi$ process.

The Gell-Mann -- Pais scheme was originally based on the assumed invariance
of the weak (decay-causing) interactions under parity (P), or mirror
reflection.  When the weak interactions were found in 1957 to violate
parity invariance, the argument was recast in terms of the product CP of
charge reflection (C) and parity, since CP
was thought at the time to be the
symmetry obeyed by the weak interactions.  The $K^0$ and $\ok$ have spin
zero.  A spin-zero final state of $\pi \pi$ necessarily has CP eigenvalue equal
to $+1$, since in the center-of-mass frame inverting space and reversing the
signs of all charges restores the original system.  Thus,
if CP is conserved, it is the CP-even linear combination of $K^0$
and $\ok$ which decays to $\pi \pi$.  With a phase convention for the neutral
kaons and the charge-conjugation operator chosen such that $CP |K^0 \rangle
= | \ok \rangle$, this is just the combination $K_1$ defined above.

Thus, in the limit of CP conservation, the neutral kaon states with
definite mass and lifetime are
\beq \label{eqn:k1k2}
K_1 = \frac{\k + \ok}{\s}~~~,~~K_2 = \frac{\k - \ok}{\s}~~~.
\eeq
These may be emulated with a pair of pendula of equal frequencies by coupling
them with a spring which also dissipates energy.  One pendulum may be thought
of as the $K^0$ and the other as the $\ok$.  In the absence of coupling, there
are two degenerate modes for which any two orthogonal linear combinations of
$K^0$ and $\ok$ can serve as an acceptable basis.  When the pendula are
coupled to one another via the spring, however, the linear combination
in which the pendula are moving with equal and opposite displacements
will be shifted in frequency and damped.  This will be the mode $K_1$ defined
above if the $K^0$ and $\ok$ are defined with appropriate phases.  The
mode in which the pendula are moving in phase with one another will
correspond to the $K_2$ and will be unshifted in frequency and undamped.
The common damping term for $K^0$ and $\ok$ corresponding to air
resistance will imply that the mode with the pendula moving in
phase with one another is also damped, but much longer-lived than the mode
which excites the spring.  Thus, damping of both eigenmodes permits
a meaningful parallel with the neutral kaon system.
A simple laboratory demonstration \cite{BW} can
be constructed which illustrates the problem up to this point.

The coupling and damping of a two-state system in quantum mechanics can be
described by a ``mass matrix'' ${\cal M}$ very similar to that discussed in
Sec.\ II A \cite{Sachs}:
\beq
i \frac{\partial}{\partial t}
\left[ \begin{array}{c} \k \\ \bk \end{array} \right]
= \m \left [ \begin{array}{c} \k \\ \bk \end{array} \right]~~~;
~~ \m = M - i \Gamma /2~~~,
\eeq
where $M$ and $\Gamma$ are Hermitian.  The eigenstates of ${\cal M}$ evolve
in proper time $t$ as $e^{- i \mu_i t}$, where $\mu_i$ ($i = 1,2$) are the
(complex) eigenvalues of ${\cal M}$.  Each $\mu_i$ may be expressed as
$\mu_i = m_i - i \gamma_i/2$, where $m_i$ is the mass of the eigenstate
$i$ and $\gamma_i$ denotes its decay width.  Here and in what follows we
are using units with $\hbar = c = 1$.  The lifetime of the eigenstate $i$
is then $\tau_i = 1/\gamma_i$.

In the CP-invariant case when the eigenstates of ${\cal M}$ are given by
(\ref{eqn:k1k2}), a little algebra shows that one must have
\beq
\m_{11} = \m_{22}~~;~~~\m_{21} = \m_{12}~~~.
\eeq
This form is only consistent with the case of the Foucault pendulum discussed
in Sec.\ II when the $\Omega$ term in Eq.~(\ref{eqn:mass}) vanishes, i.e.,
when there is no rotation.  We will see that rotation of the reference
frame in classical mechanics is analogous to CP violation in the neutral kaon
system.

When CP is violated \cite{CCFT}, the states of definite mass and lifetime $K_S$
(for ``short-lived'') and $K_L$ (for ``long-lived'') are eigenstates of a more
general mass matrix no longer obeying $\m_{21} = \m_{12}$.
However, it turns out that invariance under the product CPT, where T denotes
time reversal, still guarantees $\m_{11} = \m_{22}$.  In words,
this just says that the quantum-mechanical amplitudes for $K^0 \to K^0$ and
$\ok \to \ok$ are equal.  CPT
invariance is a very general property of quantum field theories \cite{CPT}
and we shall assume its validity here for the moment.  We shall see in the
Appendix how to relax the constraint $\m_{11} = \m_{22}$.

With CPT invariance assumed but CP violated, the eigenstates of the mass
matrix $\m$ can then be written as
\beq \label{eqn:eigS}
|S \rangle = \frac{1}{\sqrt{2(1+| \epsilon |^2)}}
\left[ (1 + \epsilon ) | \k \rangle + (1 - \epsilon )| \bk \rangle \right]~~~,
\eeq
\beq \label{eqn:eigL}
| L \rangle = \frac{1}{\sqrt{2(1+| \epsilon |^2)}}
\left[ (1 + \epsilon ) | \k \rangle - (1 - \epsilon )| \bk \rangle \right]~~~,
\eeq
where $S,~L$ denote $K_S,~K_L$, and $\epsilon$ is related to $\m$ by
\beq \label{eqn:eps}
\epsilon \equiv \frac{\sqrt{\m_{12}} - \sqrt{\m_{21}}}{\sqrt{\m_{12}} +
\sqrt{\m_{21}}} \simeq \frac{\m_{12}-\m_{21}}{4 \sqrt{\m_{12} \m_{21}}}
\simeq
\frac{{\rm Im} (\Gamma_{12}/2) + i~{\rm Im} M_{12}}{\mu_S - \mu_L}~~~.
\eeq
Here the eigenvalues $\mu_{S,L}$ of $\m$ are related to its elements by
\beq
\mu_S = \m_{11} + \sqrt{\m_{12} \m_{21}}~~; ~~~
\mu_L = \m_{11} - \sqrt{\m_{12} \m_{21}}~~~.
\eeq
The eigenstates $|S \rangle$ and $|L \rangle$ can be parametrized approximately
as 
\beq \label{eqn:appx}
|S \rangle \simeq |K_1 \rangle + \epsilon |K_2 \rangle~~~,~~
|L \rangle \simeq |K_2 \rangle + \epsilon |K_1 \rangle~~~.
\eeq
These states have a scalar product $\langle L | S \rangle \approx 2~{\rm
Re}~\epsilon$, which is no longer necessarily zero.

Both the magnitude and phase of $\epsilon$ can be measured accurately.
The squared magnitude of
$\epsilon$ is measured by comparing the rates for $K_L \to \pi \pi$ and $K_S
\to \pi \pi$; the result is  $|\epsilon| \simeq (2.28
\pm 0.02) \times 10^{-3}$.  The phase of $\epsilon$ can be measured by
experiments in which decays of $K_S$ and $K_L$ produced in a known relative
phase add coherently; the result is
${\rm Arg}~\epsilon \simeq 43^{\circ}$ \cite{PDG}.  Given what we know
about the masses and lifetimes of $K_S$ and $K_L$ \cite{PDG}, this phase agrees
with what one can derive from Eq.~(\ref{eqn:eps}) \cite{Revs}.
Specifically, one can show \cite{Revs} that $|{\rm Im} \Gamma_{12}/2| \ll
|{\rm Im} M_{12}|$.  This result then implies that
\beq
{\rm Arg}~\epsilon \approx \left\{ \begin{array}{c} 90^0 \\ 270^0 \end{array}
\right \} - ~{\rm Arg} (\mu_S - \mu_L ) ~~ {\rm for} ~~
\left \{ \begin{array}{c} {\rm Im}M_{12} > 0 \\ {\rm Im}M_{12} < 0 \end{array}
\right \}
\eeq
The observed masses $m_{S,L}$ and decay widths $\gamma_{S,L}$ in $\mu_{S,L}
= m_{S,L} - i \gamma_{S,L}/2$ are such as to imply \cite{TTTV}
${\rm Arg} ~ \epsilon = (43.5 \pm 0.1)^0$ for ${\rm Im}~ M_{12} < 0$,
and ${\rm Arg }~ \epsilon = \pi + (43.5 \pm 0.1)^0$ for ${\rm Im}~ M_{12} >
0$.

Equation~(\ref{eqn:eps}) implies that $\epsilon \ne 0$ arises from $\m_{12} \ne
\m_{21}$.  As we have seen, the Foucault pendulum with $\Omega \ne 0$ in
Eq.~(\ref{eqn:mass}) provides a mechanical illustration of this behavior.  In
the next section we shall explore this parallel a bit further.  First, we
discuss one quantity sensitive to $\epsilon$ in the neutral kaon system.  This
is the asymmetry in the semileptonic decay of a neutral kaon beam to those
final states which can arise from $K^0$ or $\bar K^0$.  We introduce this
topic because it will turn out to have a parallel in the case of the Foucault
pendulum.

We shall assume, in accord with predictions at the quark level, that 
the allowed processes are $K^0 \to \pi^- \ell^+ \nu_\ell$ and $\bar K^0 \to
\pi^+ \ell^- \bar \nu_\ell$, where $\ell = e$ or $\mu$.
Suppose, for example, that all the $K_S$ in a neutral kaon beam have decayed
away, leaving pure $K_L$.  The leptonic asymmetry
\beq
{\cal A}_\ell \equiv \frac{\Gamma(K^0 \to \pi^- \ell^+ \nu_\ell)
- \Gamma(\bar K^0 \to \pi^+ \ell^- \bar \nu_\ell)}
{\Gamma(K^0 \to \pi^- \ell^+ \nu_\ell) +
\Gamma(\bar K^0 \to \pi^+ \ell^- \bar \nu_\ell)}
\eeq
is then just ${\cal A}_\ell = 2~{\rm Re}~\epsilon$.  Measurements of this
quantity are consistent with the measured magnitude and phase of $\epsilon$.
\bigskip

\centerline{\bf IV.  PARALLELS:  BASIC PENDULUM}
\bigskip

The eigenstates $R$ and $L$ of the Foucault pendulum problem, which are
eigenvectors of (\ref{eqn:mass}), can be put into correspondence with
neutral-kaon eigenstates $S$ and $L$ by the correspondence $R \leftrightarrow
S$, $L \leftrightarrow L$.  Aside from overall phases assigned to the states,
one then sees from Eq.~(\ref{eqn:eps}) that $\epsilon = - i~{\rm sgn}(\Omega)$. 
The eigenstates have different oscillation frequencies, $\mu_S - \mu_L = 
2 \Omega$, but their lifetimes are the same.
Since $\epsilon$ is imaginary, the eigenstates $S$ and $L$ are orthogonal
to one another, just as in the CP-conserving case.

The $\pi \pi$ mode is excited by the decay of $K_1$ in (\ref{eqn:k1k2}),
which corresponds in the Foucault pendulum case to an oscillation along
the line $x = y$.  Both the eigenmodes 
(\ref{eqn:modes}) have a component along this direction.  Since the
plane of oscillation of the Foucault pendulum is always rotating (as long
as $\Omega \ne 0$), the excitation of the $\pi \pi$ mode will undergo
variations in time as this plane rotates with angular frequency $\Omega$.

The analogue of the leptonic asymmetry ${\cal A}_\ell$ discussed at the
end of the previous Section is the asymmetry of the projection of oscillations
of the $L$ eigenmode onto the $\hat x \leftrightarrow K^0$ and $\hat y
\leftrightarrow \bar K^0$ directions.  As one sees from (\ref{eqn:modes}),
there is no asymmetry, only a phase difference, with respect to oscillations in
the $\hat x$ and $\hat y$ directions.  The same conclusion can be drawn from
the fact that Re $\epsilon = 0$ in this example.
\bigskip

\centerline{\bf V.  MODIFICATIONS TO ILLUSTRATE ACTUAL KAON SYSTEM}
\bigskip

One needs eigenmodes with vastly different lifetimes in order to emulate
the true neutral-kaon system, since $\Gamma_S/\Gamma_L = \tau_L/\tau_S
\simeq 579$.  It is not difficult to simulate such eigenmodes in a
two-state system \cite{TTTV,BW}.  For example, as mentioned in Sec.\ III,
two pendula of the same
natural frequency can be coupled to one another through a dissipative spring
\cite{BW}, leading to a difference in both frequency and lifetime between the
eigenmodes in which the pendula oscillate in or out of phase with respect to
each other.

For the spherical pendulum, one can simply introduce damping in
one of the two directions, for example by using a permanent magnet as the
pendulum bob, and placing a flat coil with windings oriented along one
direction just beneath the pendulum.  The coil should be connected to
a dissipative load (e.g., a resistor).  A coil with its windings along the
$\hat x$ direction will damp oscillations in the $\hat y$ direction.  However,
if we were to treat the $\hat x$ and $\hat y$ damping differently,
we would be violating CPT invariance, since then $\m_{22} \ne \m_{11}$.

In the present Section we wish to emulate a CPT-preserving system, which
is most convenient in a new basis.  In the $(K^0,\bar K^0)$
basis a CPT-preserving mass matrix has the form
\beq
\m = \left[ \begin{array}{c c} \m_{11} & \m_{12} \\ \m_{21} & \m_{11}
\end{array} \right]~~~.
\eeq
Thus one must have equal natural frequencies and damping terms for oscillations
in the $\hat x$ and $\hat y$ directions.  However, one can transform \cite{SW}
to the basis $(K_1,K_2)$ corresponding to oscillations in the $(\k \pm \bk)/\s
= (\hat x \pm \hat y)/\s$
directions.  In this basis the mass matrix is $\n = U \m U^\dag$, where
\beq
U \equiv \frac{1}{\s} \left[ \begin{array}{c c} 1 & 1 \\ 1 & -1 \end{array}
\right] = U^\dag~~~.
\eeq
One finds
\beq
\n = \left[ \begin{array}{c c} \n_{11} & \n_{12} \\ - \n_{12} & \n_{22}
\end{array} \right]~~~,
\eeq
where $\n_{11} = \m_{11} + (\m_{12} + \m_{21})/2$, $N_{22} = \m_{11}
- (\m_{12} + \m_{21})/2$, and $\n_{12} = (\m_{21} - \m_{12})/2$.  Thus,
in this basis, one can have a mass matrix of the form
\beq
\n = \left[ \begin{array}{c c} \omega_1 - i \gamma_1 & i \Omega' \\
- i \Omega' & \omega_2 - i \gamma_2 \end{array} \right]~~~,
\eeq
where $\Omega'$ is not necessarily real.  The CPT-invariance is expressed
in the $K_1$--$K_2$ basis by the condition $\n_{21} = - \n_{12}$.

Physically one can emulate the system described by the matrix $\n$ by having a
Foucault pendulum with different damping in the $K_1$ and $K_2$
directions.  Such damping could be implemented, for example, by the
inductive setup noted above.  Natural frequencies in two orthogonal directions
can be made to differ using a hinged set of supports. One could also introduce
different damping constants in two perpendicular directions through properties
of the hinged joints themselves.

The Foucault pendulum case corresponds to real $\Omega'$ for the new basis.
Note that $\Omega'$ in $\n$ is then the same as $\Omega$ in $\m$;
aside from a sign flip in the off-diagonal elements, $\n$ and $\m$ are the
same matrix.

For $\omega_1 - i \gamma_1 \ne \omega_2 - i \gamma_2$ and $\Omega \ll
\omega_{1,2}$, the eigenvectors and their corresponding eigenvalues are
approximately
\beq
|S \rangle = \left[ \begin{array}{c} 1 \\ \epsilon_S \end{array} \right]~~,~~~
\mu_S = \omega_1 - i \gamma_1 + \delta_S~~~,
\eeq
\beq
\epsilon_S = \frac{- i \Omega}{\omega_1 - \omega_2 - i(\gamma_1 - \gamma_2)}~~,
~~~\delta_S = \frac{\Omega^2}{\omega_1 - \omega_2 - i(\gamma_1 - \gamma_2)}~~~,
\eeq
\beq
|L \rangle = \left[ \begin{array}{c} \epsilon_L \\ 1 \end{array} \right]~~,~~~
\mu_L = \omega_2 - i \gamma_2 + \delta_L~~~,
\eeq
\beq
\epsilon_L = \frac{- i \Omega}{\omega_1 - \omega_2 - i(\gamma_1 - \gamma_2)} =
\epsilon_S \equiv \epsilon~~,~~~
\delta_L = - i \Omega \epsilon = - \delta_S~~~.
\eeq

Let us now discuss the time-evolution of states which are initially
$|S \rangle$ and $|L \rangle$:
\beq
|S \rangle = \evS \to \evS e^{-i \mu_S t}~~,~~~
|L \rangle = \evL \to \evL e^{-i \mu_L t}~~~.
\eeq
Defining $\phi \equiv {\rm Arg}~\epsilon$, these lead to the time-dependences
\beq
{\rm Re}~\vx_S(t) = x_0{\rm Re} \evS e^{-i \mu_S t} \simeq e^{-\gamma_1 t}
\left[ \begin{array}{c} \cos \omega_1 t \\ |\epsilon| \cos (\omega_1 t - \phi)
\end{array} \right]~~~,
\eeq
\beq
{\rm Re}~\vx_L(t) = x_0{\rm Re} \evL e^{-i \mu_L t} \simeq e^{-\gamma_2 t}
\left[ \begin{array}{c} |\epsilon| \cos (\omega_2 t - \phi) \\ \cos \omega_2 t
\end{array} \right]~~~'
\eeq
where contributions of order $\delta_{S,L}$ have been omitted.
Thus both $S$ and $L$ eigenstates correspond to orbits which are slightly
rotated to favor an orientation in the direction of $K^0$-like
decays by an amount proportional to $\epsilon$.  The Coriolis force in
this example induces an effect which is akin to the leptonic asymmetry
parameter ${\cal A}_\ell$ discussed at the end of Sec.~III.  Moreover, the
decay to $\pi \pi$ in the $K_1$--$K_2$ basis is like the x-projection of the
eigenstate.  Although the dominant damping is along the $\hat x$-direction
(we assume $\gamma_1 \gg \gamma_2$), the presence of the Coriolis force causes
the eigenstate $L$ to have a projection proportional to $\epsilon$
along $\hat x$, i.e., the $K_L$ does decay to $\pi \pi$.
\bigskip

\centerline{\bf VI.  CONCLUSIONS}
\bigskip

We have shown that the motion of a Foucault pendulum has many features
in common with the CP-violating neutral kaon system.  When the natural
frequencies for oscillation in two perpendicular directions and the
damping terms for these directions are equal, the parameter $\epsilon$
describing the eigenstates $|K_S \rangle \simeq |K_1 \rangle + \epsilon
|K_2 \rangle$ and $|K_L \rangle \simeq |K_2 \rangle + \epsilon |K_1 \rangle$
takes on the special value $\epsilon = - i$.  In order to simulate the
physical situation in which $|\epsilon| = {\cal O}(2 \times 10^{-3})$
and Arg$(\epsilon) \simeq \pi/4$, one must construct a Foucault pendulum
with slightly different natural frequencies in two perpendicular directions,
and with vastly different damping constants in these directions.  While the
practical realization of such a construction sounds challenging, it is
interesting that, at least in principle, it seems feasible entirely within
the realm of classical mechanics.

As we show in the Appendix, the phenomenon of CPT violation, with preservation
of T and violation of CP,
can be emulated by coupled resonant circuits, building upon the results of
Ref.~\cite{TTTV}.  The program set forth in that work is still incomplete;
we would be delighted to see an implementation of CP and T violation, with
CPT conservation, through the asymmetric coupling of two resonant circuits with
equal frequencies.

The classical emulation of CP violation via the Foucault pendulum leaves us
with one big puzzle.  In the classical system, the asymmetry in coupling
between the two modes is imposed from the outside, so to speak, through the
Earth's rotation.  In the neutral-kaon system, the corresponding asymmetry
in $\k$--$\bk$ mixing is thought to arise from a complex phase in the
Cabibbo-Kobayashi-Maskawa matrix \cite{CKM} describing the charge-changing
weak transitions of quarks.  Is that phase an indication of a new fundamental
asymmetry arising from physics beyond the Standard Model?
\bigskip

\centerline{\bf ACKNOWLEDGMENTS}
\bigskip

We thank Bruno Carneiro da Cunha for the original suggestion which led to
this investigation.  This work was supported in part by the
United States Department of Energy under Grant No. DE FG02 90ER40560. 
\bigskip

\centerline{\bf APPENDIX.  CPT-VIOLATING CASE}
\bigskip

The matrices $\m$ and $\n$ are arbitrary when CPT is violated.  The relation
between them is
$$
\n_{11} = (\m_{11} + \m_{12} + \m_{21} + \m_{22})/2~~,~~~
\n_{12} = (\m_{11} - \m_{12} + \m_{21} - \m_{22})/2~~~,
$$
\beq
\n_{21} = (\m_{11} + \m_{12} - \m_{21} - \m_{22})/2~~,~~~
\n_{22} = (\m_{11} - \m_{12} - \m_{21} + \m_{22})/2~~~.
\eeq
A simple example of a CPT-violating mass matrix $\n$ involves coupling
between two resonant circuits, as discussed in Ref.~\cite{TTTV}.  If the
circuits are taken to have different frequencies, the matrix takes the form
\beq
\n = \left[ \begin{array}{c c} \omega_1 - i \gamma_1 & \alpha \\
\alpha & \omega_2 - i \gamma_2 \end{array} \right]~~~,
\eeq
in which the off-diagonal elements are equal (rather than equal and opposite
as in the CPT-preserving case).  An analysis parallel to that for the
eigenstates $S$ and $L$ performed in the previous Section leads to the
results (for $|\alpha| \ll \omega_{1,2}$)
\beq
|S \rangle = \left[ \begin{array}{c} 1 \\ \epsilon_S \end{array} \right]~~,~~~
|L \rangle = \left[ \begin{array}{c} \epsilon_L \\ 1 \end{array} \right]~~~,
\eeq
with
\beq
\epsilon_S = \frac{\alpha}{\omega_1 - \omega_2 - i(\gamma_1 - \gamma_2)}
= - \epsilon_L~~~ \equiv \te,
\eeq
\beq
\mu_S = \omega_1 - i \gamma_1 + \frac{\alpha^2}{\omega_1 - \omega_2
- i(\gamma_1 - \gamma_2)}~~,~~~
\mu_L = \omega_2 - i \gamma_2 - \frac{\alpha^2}{\omega_1 - \omega_2
- i(\gamma_1 - \gamma_2)}~~~.
\eeq
The eigenstates are thus
\beq \label{eqn:te}
|S \rangle \simeq |1 \rangle + \te |2 \rangle~~,~~~
|L \rangle \simeq |2 \rangle - \te |1 \rangle~~~.
\eeq
This case (see, e.g., Ref.~\cite{Dalitz}) correponds to invariance with respect
to time-reversal, so that CP and CPT are violated.  Both (\ref{eqn:appx})
and (\ref{eqn:te}) can be written in the more general form
\beq
|S \rangle \simeq |1 \rangle + \epsilon_S |2 \rangle~~,~~~
|L \rangle \simeq |2 \rangle + \epsilon_L |1 \rangle~~~.
\eeq
The CPT-preserving, CP-violating case corresponds to $\epsilon_S = \epsilon_L
= \epsilon$, while the case (\ref{eqn:te}) corresponds to $\epsilon_S =
- \epsilon_L = \te$.  To lowest order in $\epsilon_{S,L}$, one finds
\beq
|\k \rangle = \frac{1}{\s}[|S \rangle (1-\epsilon_L) + |L \rangle
(1-\epsilon_S)]~~,~~~
|\bk \rangle = \frac{1}{\s}[|S \rangle (1+\epsilon_L) - |L \rangle
(1 + \epsilon_S)]~~~.
\eeq 
Since the states $|S,L \rangle$ evolve with proper time $t$ as
$|S,L \rangle \to e^{-i \mu_{S,L} t} |S,L \rangle$, a short calculation
shows that
$$
|\k \rangle \to |\k \rangle [f_+(t) + (\epsilon_S - \epsilon_L) f_-(t)]
+ |\bk \rangle [1 -(\epsilon_S + \epsilon_L)]f_-(t)~~~,
$$
\beq
|\bk \rangle \to |\bk \rangle [ f_+(t) + (\epsilon_L - \epsilon_S)f_-(t)]
+ |\k \rangle [1 + (\epsilon_S + \epsilon_L)]f_-(t)~~~,
\eeq
where $f_{\pm}(t) \equiv [\exp(-i \mu_S t) \pm \exp(-i \mu_L t)]/2$.

For $\epsilon_L = - \epsilon_S$, the evolution of $\k$ into $\bk$ is the
same as that for $\bk$ into $\k$, corresponding to a time-reversal-invariant
situation.  However, the amplitudes for $\k \to \k$ and $\bk \to \bk$
differ from one another, corresponding to CPT violation.

For $\epsilon_L = \epsilon_S = \epsilon$, the amplitudes for $\k \to \k$
and $\bk \to \bk$ are the same, corresponding to $CPT$ invariance, while
those for $\k \to \bk$ and $\bk \to \k$ differ from one another, corresponding
to T violation.  In this case the terms $|f_-(t)|^2$ cancel in the rate
asymmetry
\beq
{\cal A}_T \equiv \frac{\Gamma[\k(0) \to \ok(t)] - \Gamma[\ok(0) \to \k(t)]}
{\Gamma[\k(0) \to \ok(t)] + \Gamma[\ok(0) \to \k(t)]}~~~,
\eeq
and to lowest order one finds \cite{KabirT} ${\cal A}_T = 4~{\rm Re}~\epsilon$.
This relation has recently passed an experimental test at CPLEAR \cite{CPLEAR}.
\bigskip

\def \ajp#1#2#3{Am. J. Phys. {\bf#1}, #2 (#3)}
\def \apny#1#2#3{Ann. Phys. (N.Y.) {\bf#1}, #2 (#3)}
\def \app#1#2#3{Acta Phys. Polonica {\bf#1}, #2 (#3)}
\def \arnps#1#2#3{Ann. Rev. Nucl. Part. Sci. {\bf#1}, #2 (#3)}
\def \cmts#1#2#3{Comments on Nucl. Part. Phys. {\bf#1}, #2 (#3)}
\def \cn{Collaboration}
\def \cp89{{\it CP Violation,} edited by C. Jarlskog (World Scientific,
Singapore, 1989)}
\def \efi{Enrico Fermi Institute Report No. EFI}
\def \epjc#1#2#3{Eur.~Phys.~J.~C {\bf#1}, #2 (#3)}
\def \f79{{\it Proceedings of the 1979 International Symposium on Lepton and
Photon Interactions at High Energies,} Fermilab, August 23-29, 1979, ed. by
T. B. W. Kirk and H. D. I. Abarbanel (Fermi National Accelerator Laboratory,
Batavia, IL, 1979}
\def \hb87{{\it Proceeding of the 1987 International Symposium on Lepton and
Photon Interactions at High Energies,} Hamburg, 1987, ed. by W. Bartel
and R. R\"uckl (Nucl. Phys. B, Proc. Suppl., vol. 3) (North-Holland,
Amsterdam, 1988)}
\def \ib{{\it ibid.}~}
\def \ibj#1#2#3{~{\bf#1}, #2 (#3)}
\def \ichep72{{\it Proceedings of the XVI International Conference on High
Energy Physics}, Chicago and Batavia, Illinois, Sept. 6 -- 13, 1972,
edited by J. D. Jackson, A. Roberts, and R. Donaldson (Fermilab, Batavia,
IL, 1972)}
\def \ijmpa#1#2#3{Int. J. Mod. Phys. A {\bf#1}, #2 (#3)}
\def \ite{{\it et al.}}
\def \jpb#1#2#3{J.~Phys.~B~{\bf#1}, #2 (#3)}
\def \lkl87{{\it Selected Topics in Electroweak Interactions} (Proceedings of
the Second Lake Louise Institute on New Frontiers in Particle Physics, 15 --
21 February, 1987), edited by J. M. Cameron \ite~(World Scientific, Singapore,
1987)}
\def \kdvs#1#2#3{{Kong.~Danske Vid.~Selsk., Matt-fys.~Medd.} {\bf #1}, No.~#2
(#3)}
\def \ky85{{\it Proceedings of the International Symposium on Lepton and
Photon Interactions at High Energy,} Kyoto, Aug.~19-24, 1985, edited by M.
Konuma and K. Takahashi (Kyoto Univ., Kyoto, 1985)}
\def \mpla#1#2#3{Mod. Phys. Lett. A {\bf#1}, #2 (#3)}
\def \nat#1#2#3{Nature {\bf#1}, #2 (#3)}
\def \nc#1#2#3{Nuovo Cim. {\bf#1}, #2 (#3)}
\def \np#1#2#3{Nucl. Phys. {\bf#1}, #2 (#3)}
\def \PDG{Particle Data Group, L. Montanet \ite, \prd{50}{1174}{1994}}
\def \pisma#1#2#3#4{Pis'ma Zh. Eksp. Teor. Fiz. {\bf#1}, #2 (#3) [JETP Lett.
{\bf#1}, #4 (#3)]}
\def \pl#1#2#3{Phys. Lett. {\bf#1}, #2 (#3)}
\def \pla#1#2#3{Phys. Lett. A {\bf#1}, #2 (#3)}
\def \plb#1#2#3{Phys. Lett. B {\bf#1}, #2 (#3)}
\def \pr#1#2#3{Phys. Rev. {\bf#1}, #2 (#3)}
\def \prc#1#2#3{Phys. Rev. C {\bf#1}, #2 (#3)}
\def \prd#1#2#3{Phys. Rev. D {\bf#1}, #2 (#3)}
\def \prl#1#2#3{Phys. Rev. Lett. {\bf#1}, #2 (#3)}
\def \prp#1#2#3{Phys. Rep. {\bf#1}, #2 (#3)}
\def \ptp#1#2#3{Prog. Theor. Phys. {\bf#1}, #2 (#3)}
\def \rmp#1#2#3{Rev. Mod. Phys. {\bf#1}, #2 (#3)}
\def \rp#1{~~~~~\ldots\ldots{\rm rp~}{#1}~~~~~}
\def \si90{25th International Conference on High Energy Physics, Singapore,
Aug. 2-8, 1990}
\def \slc87{{\it Proceedings of the Salt Lake City Meeting} (Division of
Particles and Fields, American Physical Society, Salt Lake City, Utah, 1987),
ed. by C. DeTar and J. S. Ball (World Scientific, Singapore, 1987)}
\def \slac89{{\it Proceedings of the XIVth International Symposium on
Lepton and Photon Interactions,} Stanford, California, 1989, edited by M.
Riordan (World Scientific, Singapore, 1990)}
\def \smass82{{\it Proceedings of the 1982 DPF Summer Study on Elementary
Particle Physics and Future Facilities}, Snowmass, Colorado, edited by R.
Donaldson, R. Gustafson, and F. Paige (World Scientific, Singapore, 1982)}
\def \smass90{{\it Research Directions for the Decade} (Proceedings of the
1990 Summer Study on High Energy Physics, June 25--July 13, Snowmass, Colorado),
edited by E. L. Berger (World Scientific, Singapore, 1992)}
\def \tasi90{{\it Testing the Standard Model} (Proceedings of the 1990
Theoretical Advanced Study Institute in Elementary Particle Physics, Boulder,
Colorado, 3--27 June, 1990), edited by M. Cveti\v{c} and P. Langacker
(World Scientific, Singapore, 1991)}
\def \yaf#1#2#3#4{Yad. Fiz. {\bf#1}, #2 (#3) [Sov. J. Nucl. Phys. {\bf #1},
#4 (#3)]}
\def \zhetf#1#2#3#4#5#6{Zh. Eksp. Teor. Fiz. {\bf #1}, #2 (#3) [Sov. Phys. -
JETP {\bf #4}, #5 (#6)]}
\def \zpc#1#2#3{Zeit. Phys. C {\bf#1}, #2 (#3)}
\def \zpd#1#2#3{Zeit. Phys. D {\bf#1}, #2 (#3)}

\end{document}